\documentclass[prl,aps,twocolumn,amsmath,superscriptaddress,showpacs]{revtex4} 
\usepackage{graphicx}

\begin{document}
\title{Stochastic unraveling of time-local quantum master equations \\
  beyond the Lindblad class}
\author{Ulrich Kleinekath\"ofer}
\affiliation{International University Bremen, P.O.Box 750 561, 28725
  Bremen, Germany}
\affiliation{Institut f\"ur Physik, Technische Universit\"at, 09107 Chemnitz,
 Germany}
\author{Ivan Kondov}
\affiliation{Institut f\"ur Physik, Technische Universit\"at, 09107 Chemnitz,
 Germany}
\author{Michael Schreiber}
\affiliation{International University Bremen, P.O.Box 750 561, 28725
  Bremen, Germany}
\affiliation{Institut f\"ur Physik, Technische Universit\"at, 09107 Chemnitz,
 Germany}
\date{ \today}
\begin{abstract}
  A new method for stochastic unraveling of general time-local quantum
  master equations (QME) which involve the reduced density operator at time
  $t$ only is proposed. The present kind of jump algorithm enables a
  numerically efficient treatment of QMEs which are not of Lindblad form.
  So it opens new large fields of application for stochastic methods.  The
  unraveling can be achieved by allowing for trajectories with negative
  weight. We present results for the quantum Brownian motion and the Redfield
  QMEs as test examples. The algorithm can also unravel non-Markovian QMEs
  when they are in a time-local form like in the time-convolutionless
  formalism.
\end{abstract}


\maketitle


Quantum master equations (QMEs) are frequently used to describe
time-independent as well as time-dependent phenomena in chemical physics,
quantum optics, solid state physics, biological physics, etc.\ (see
Ref.~\onlinecite{weis99} for a number of typical examples).  These QMEs
describe the time evolution of density matrices which are used in order to
represent the mixed nature of the states.  Stochastic unraveling is an
efficient numerical tool for solving such equations. This method allows one
to simulate much larger and more complex systems with many degrees of
freedom.  It can, for example, be used to accurately describe femtochemical
experiments in the liquid phase whose description has been limited until
now to models with one or two effective interaction coordinates.  In the
unraveling scheme one considers an ensemble of stochastic Schr\"odinger
equations (SSEs) which in the limit of a large ensemble resembles the
respective QME.  The numerical effort scales much more favorably with the
size of the basis since one is now dealing with wave functions and not
density matrices anymore 
(for a comparison of direct integrators, see Ref.~\onlinecite{kond01}).
 Another aspect of the stochastic methods is the
possible physical interpretation of experiments detecting macroscopic
fluctuations (e.g. photon counting) in various quantum systems
\cite{plen98}.  Most of the unraveling schemes
\cite{dali92,gard92,gisi92,garr94,wolf95,plen98} have been restricted to
QMEs of Lindblad form \cite{lind75} which ensures that the reduced density
matrix (RDM) stays positive semi-definite for all times and all parameters.
Nevertheless there are many physical meaningful QMEs which result in
positive- or almost positive-definite RDMs although they are not of
Lindblad form.  The increasing interest in descriptions beyond the Lindblad
class such as the quantum Brownian motion \cite{stru99,yu99}, the Redfield
formalism \cite{may00}, non-Markovian schemes \cite{cape94b,meie99,breu99},
etc.\ resulted in various efforts to develop new stochastic wave function
algorithms.

Strunz et al.~\cite{stru99,yu99} developed the non-Markovian Quantum
Diffusion Model. In general, this method can also be applied to QMEs in
Markov approximation even though they might not preserve positivity (see
also Ref.~\cite{budi00}).  A similar approach was also proposed by Gaspard
et al.~\cite{gasp99b}.  Very recently Stockburger and Grabert\cite{stoc02}
developed a method how to exactly represent the RDM of a system coupled to
a linear heat bath in terms of SSEs. The numerical properties of this
approach need to be explored.  Breuer et al.~\cite{breu99} extended a
scheme which they had used to calculate multi-time correlation functions
\cite{breu97} to the unraveling of QMEs.  Their technique is based on
doubling the Hilbert space. Instead of a single stochastic wave function
one has a pair of them \cite{breu99}.  This scheme conserves Hermiticity of
the RDM only on average and not for every single realization.  Thus, the
deviation from Hermiticity is a quantity with statistical error and one has
to perform a huge number of realizations in order to achieve good
convergence.  Since stability and efficiency are crucial issues for
unraveling algorithms we propose in this article an alternative approach
which fulfills these criteria.

The aim is to represent, in terms of quantum trajectories, the solution
$\rho(t)$ of a generalized time-local Hermiticity-conserving QME
\begin{eqnarray}
\label{eq:qme}
\frac{d\rho(t)}{dt} &=& A(t)\rho(t)+\rho(t)A^\dagger(t) \nonumber \\
&&+\sum\limits_{k=1}^M \big\{C_k(t)\rho(t) E_k^\dagger(t)
+E_k(t)\rho(t)C_k^\dagger(t)\big\}
\end{eqnarray}
with the total number $M$ of dissipative channels and arbitrary operators
$A(t)$, $C_k(t)$, and $E_k(t)$. Examples for these operatores are given
below.  Here we restrict the operators in such a way that the norm of the
solution stays conserved. For readability we shall omit the time arguments
in the following.

In order to approach the problem let us define a state vector
$\left(|\psi\rangle,|\phi\rangle\right)^T$ spanning a doubled Hilbert space
as proposed in Ref.~\onlinecite{breu99}. Unlike Ref.~\onlinecite{breu99}
the RDM shall be reproduced by an ensemble average (denoted by overbars) of
outer products of the vectors $|\psi\rangle$ and $|\phi\rangle$:
\begin{equation}
\rho = \overline{
 |\psi\rangle\langle\phi|}+\overline{|\phi\rangle\langle\psi|}~.
\label{eq:recon}
\end{equation}
A particular realization of the stochastic process will be denoted by the
pair $\left(|\psi\rangle,|\phi\rangle\right)$. The averaging is performed
over all trajectories possibly including a weighted sum over pure initial
states. A vantage of this averaging is the conservation of Hermiticity
for every single trajectory in contrast to Ref.~\cite{breu99}. We note that
this small modification improves the numerical efficiency significantly.

For the SSEs let us consider $2M$ independent possibly complex noise
variables $\xi^i_k(t)$. The superscripts denote which of the two terms from
the Hermitian pair in the sum in Eq.~(\ref{eq:qme}) is taken and subscripts
denote the various dissipative channels. All stochastic differentials
$d\xi^i_k(t)$ are assumed to have zero mean, to be normalized and
uncorrelated \cite{gard85}:
\begin{eqnarray}
\overline{d\xi^i_k}=0,\ \overline{d\xi^{i\ast}_k d\xi^j_{l}}=
\delta_{ij}\delta_{kl}dt~.
\label{eq:noise}
\end{eqnarray}
Next, as an {\it ansatz} we construct a SSE which propagates
the pair $\left(|\psi\rangle,|\phi\rangle\right)$
\begin{subequations}
\label{eq:sse0}
\begin{eqnarray}
d|\psi\rangle&=&
D_1|\psi\rangle dt
+\sum\limits_{k=1}^M\sum\limits_{i=1}^2
 S_{1k}^i|\psi\rangle d \xi_k^i~,
\label{eq:sse0_a}
\\
d|\phi\rangle&=&
D_2|\phi\rangle dt
+\sum\limits_{k=1}^M\sum\limits_{i=1}^2
 S_{2k}^i|\phi\rangle d \xi_k^i~.
\label{eq:sse0_b}
\end{eqnarray}
\end{subequations}
The operators $D_1$ and $D_2$ govern the deterministic and the operators
$S_{jk}^i$ the stochastic part of the evolution. In general, they may
depend on the state vector and explicitly on time. After differentiating
Eq.~(\ref{eq:recon}), neglecting all terms higher than first order in $dt$,
and assuming that ensemble averages always factorize \cite{dors00a} one
obtains
\begin{eqnarray}
d\rho&=&
\left[
 D_1 \overline{|\psi\rangle\langle\phi|}
+D_2 \overline{|\phi\rangle\langle\psi|}
\right] dt
\\
&&+\sum\limits_{k=1}^M
\left[
 S_{1k}^1|\overline{\psi\rangle\langle\phi|}S_{2k}^{1\dagger}
+S_{2k}^2|\overline{\phi\rangle\langle\psi|}S_{1k}^{2\dagger}
\right] dt
+\ h.c.  \nonumber
\label{eq:qme_sub}
\end{eqnarray}
Comparing with Eq.~(\ref{eq:qme}) one notes that $S_{1k}^1$
has to equal $S_{2k}^2$ and $S_{2k}^1$ has to equal $S_{1k}^2$.
Moreover, one can see that $S_{2k}^1=C_k+\alpha_k^1$ and
$S_{2k}^2=E_k+\alpha_k^2$ with $\alpha_k^1$ and $\alpha_k^2$ being
arbitrary scalar functions of $\left(|\psi\rangle,|\phi\rangle\right)^T$
and possibly of time. Making the latter substitutions in
Eq.~(\ref{eq:qme_sub}) yields the constraint
\begin{eqnarray}
D_1=D_2=A-\sum\limits_{k=1}^M
\left(
\alpha_k^{2\ast} C_k +\alpha_k^{1\ast} E_k + \alpha_k^1\alpha_k^{2\ast}
\right)~.
\label{eq:d-constr}
\end{eqnarray}

Any quantum jump method is specified by jump rates $p^i_k$ which have to be
real scalar functions of $\left(|\psi\rangle,|\phi\rangle\right)$. If
$n_k^i(t)$ is the number of jumps in channel $k$ and due to term $i$
up to time $t$, the probability for $n_k^i(t)$ to increase by one, i.e. the
expectation value of both $dn_k^i$ and $(dn_k^i)^2$, is equal to $p^i_k dt$
during the infinitesimal time interval $dt$. Thus, the noise variables
$\xi^i_k$ obeying  Eq.~(\ref{eq:noise}) are related to $dn_k^i(t)$ as
\cite{dors00a}
\begin{eqnarray}
d\xi^i_k=\frac{dn^i_k - p^i_k dt}{\sqrt{p^i_k}}e^{i\varphi}~.
\label{eq:noise-jumprate}
\end{eqnarray}
The phase factor $e^{i\varphi}$ does not 
change the RDM expressions within each realization and can be set to one.
Substituting Eq.~(\ref{eq:noise-jumprate}) into Eq.~(\ref{eq:sse0}) one
finds that $\alpha_k^i=-\sqrt{p^i_k}$. So the SSEs for our quantum jump
method  read
\begin{subequations}
\label{eq:sse1}
\begin{eqnarray}
\label{eq:sse1_a}
d|\psi\rangle&=&
\left(
A+\sum\limits_{k=1}^M\frac{p_k^1+p_k^2}{2}
\right)
|\psi\rangle dt
 \\
&&+\sum\limits_{k=1}^M
\left[ 
  \left( \frac{E_k}{\sqrt{p_k^1}}-1 \right) dn_k^1
+ \left( \frac{C_k}{\sqrt{p_k^2}}-1 \right) dn_k^2
\right]
|\psi\rangle, \nonumber
\\
\label{eq:sse1_b}
d|\phi\rangle&=&
\left(
A+\sum\limits_{k=1}^M\frac{p_k^1+p_k^2}{2}
\right)
|\phi\rangle dt
 \\
&&+\sum\limits_{k=1}^M
\left[ 
  \left( \frac{C_k}{\sqrt{p_k^1}}-1 \right) dn_k^1
+ \left( \frac{E_k}{\sqrt{p_k^2}}-1 \right) dn_k^2
\right]
|\phi\rangle. \nonumber
\end{eqnarray}
\end{subequations}

The jump rates $p_k^1$ and $p_k^2$ still remain free parameters. In the
statistical limit their values have no influence on any averaged physical
quantity. Nevertheless, it turns out that they can strongly influence the
convergence behavior of the jump algorithm, i.e. they determine the
statistical error of the observables calculated. A detailed discussion of
this influence and utilization of such free parameters can be found in
Ref.~\onlinecite{felb99}.

To ensure an efficient scheme with fast convergence one has to 
require that the norm of every single trajectory is constant in time.
Asking for $\langle \phi|\phi \rangle$, $\langle \psi|\psi \rangle$, etc.\
being constant in time
does not create a stable scheme but the condition of norm preservation of
$|\psi\rangle\langle\phi|+|\phi\rangle\langle\psi|$
\begin{eqnarray}
{\rm Tr}\left\{\frac{d}{dt}\left[|\psi\rangle\langle\phi|+
|\phi\rangle\langle\psi|\right]\right\}=0
\label{eq:nconserve}
\end{eqnarray}
does. Unfortunately, applying this condition does not lead to positive
values of the jump rates $p_k^i$ for all trajectories at all times.
However, since the $p_k^i$ are arbitrary real functions, they can be
replaced by their absolute values.  The price to pay is that we have to
introduce an additional weight factor for the trajectories which jumps
between one and minus one. In addition, there is a small deviation of the
norm from unity because in the regions where the $p_k^i$ are replaced by
their absolute values norm conservation is no longer guaranteed.  But in
all our tests this deviation was far below one percent and neither effected
numerical stability nor efficiency.  The negative weights are actually
needed to reconstruct RDMs which are, in general, not
positive-semi-definite.  If the RDM stays positive-semi-definite during its
entire time-evolution the negative weights of some trajectories are not
needed, i.e.\ all trajectories can be normalized to unity and represent
physically pure states of the open quantum system.  In the examples below
the RDM can exhibit negative populations.  This unphysical situation could
probably be cured by applying an initial slippage to the initial state
\cite{gasp99a,yu00}. We note that this physically unreasonable RDMs occur
because of unphysical initial states or because the QME is not physically
correct or is applied in a parameter region where it is is not valid.
Nevertheless an unraveling scheme has to be able to mimic also this
unphysical behavior of the QME because in the ensemble average both should
fully coincide.

The condition (\ref{eq:nconserve}) applied to the QME~(\ref{eq:qme})
results in the additional constraint
\begin{eqnarray}
\label{eq:a-c-e}
A+A^\dagger=-\sum_{k=1}^M
\left(
 E^\dagger_k C^{\phantom\dagger}_k
+C^\dagger_k E^{\phantom\dagger}_k
\right)
\end{eqnarray}
and if applied to the deterministic part of the corresponding
SSE~(\ref{eq:sse1}) yields the total jump rate
\begin{eqnarray}
\label{eq:tot-rate}
p=-\frac
{\langle\phi|A+A^\dagger|\psi\rangle
+\langle\psi|A+A^\dagger|\phi\rangle}
{\langle\phi|\psi\rangle+\langle\psi|\phi\rangle}.
\end{eqnarray}
All partial jump rates can be found subsequently making use of
Eqs.~(\ref{eq:a-c-e}) and (\ref{eq:tot-rate}):
\begin{subequations}
\label{eq:jump-rates}
\begin{eqnarray}
p_k^1&=&\frac{
 \langle\phi|C_k^\dagger E^{\phantom\dagger}_k|\psi\rangle
+\langle\psi|E_k^\dagger C^{\phantom\dagger}_k|\phi\rangle}
{\langle\phi|\psi\rangle+\langle\psi|\phi\rangle},
\label{eq:jump-rates_a}
\\
p_k^2&=&\frac{
 \langle\phi|E_k^\dagger C^{\phantom\dagger}_k|\psi\rangle
+\langle\psi|C_k^\dagger E^{\phantom\dagger}_k|\phi\rangle}
{\langle\phi|\psi\rangle+\langle\psi|\phi\rangle}.
\label{eq:jump-rates_b}
\end{eqnarray}
\end{subequations}

\begin{figure}
\includegraphics[width=8.2cm]{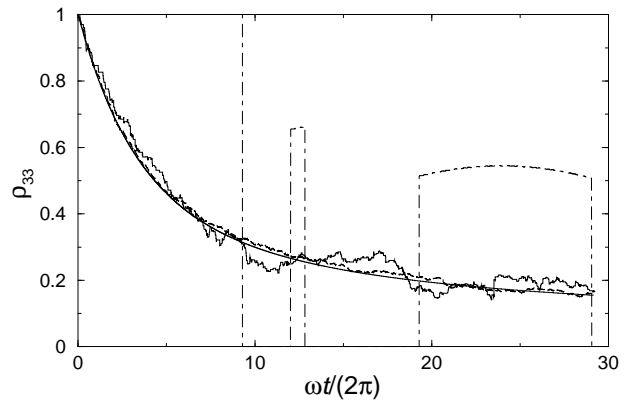}
\caption{\label{fig:qbm}
Time evolution of the third excited state of the harmonic oscillator in the
quantum Brownian oscillator model for $\gamma=10^{-3}\omega$,
$kT=4.5\omega$.  The direct integration of the QME (thick solid line) is
compared to the results of the quantum jump method with 1 trajectory
(dot-dashed line), average of 100 (thin solid line) and 1000 (broken line)
trajectories.}
\end{figure}

In the rest of this article let us briefly show how the proposed method can
be applied to two typical physical problems: the quantum Brownian motion
and dissipative electron transfer within Redfield theory. In both cases the
systems are described by Markovian QMEs which do not have Lindblad
structure. The model of Brownian motion \cite{weis99} describes a particle
with mass $m$, coordinate $q$, momentum $p$ and Hamiltonian $H_{\rm S}$
interacting with a thermal bath.  In the high temperature limit of a bath
of harmonic oscillators the relevant QME has the form
\begin{eqnarray}
\label{eq:qbm}
\frac{d\rho}{dt}=-\frac{i}{\hbar} 
\left[
H_{\rm S},\rho
\right]
-\frac{i\gamma}{2\hbar}
\left[q,
\{p,\rho\}
\right]
-\frac{m\gamma k T}{\hbar^2}
\left[
q,\left[q,\rho\right]
\right]
\end{eqnarray}
where $\gamma$ is the damping rate. Comparing with
Eq.~(\ref{eq:qme}) one finds the operators of the jump algorithm ($M=2$)
\begin{subequations}
\label{eq:qbm-oper}
\begin{eqnarray}
E_1&=&\sqrt{\frac{\gamma}{2\hbar}}q,
\ 
C_1=-i\sqrt{\frac{\gamma}{2\hbar}}p,
\label{eq:qbm-oper_a}
\\
E_2&=&\sqrt{\frac{m\gamma k T}{\hbar^2}}q,
\ 
C_2=E_2,
\label{eq:qbm-oper_b}
\\
A&=&-\frac{i}{\hbar}H_{\rm S}+\frac{i\gamma}{2\hbar}q p-\frac{m\gamma k T}{\hbar^2}q q.
\label{eq:qbm-oper_c}
\end{eqnarray}
\end{subequations}
Modeling the particle as a harmonic oscillator with eigenfrequency $\omega$
one can compute the population dynamics depicted in Fig.~\ref{fig:qbm}.
The initial state of the oscillator is the pure state $\rho_{33}=1$. As can
be seen, the agreement of the results using our new stochastic method and
using a direct integration of the QME is already quite good for one
thousand samples.

\begin{figure}
\includegraphics[width=8.cm]{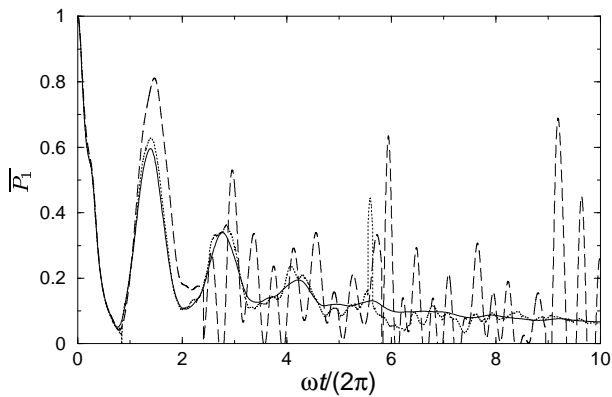}
\caption{
\label{fig:et-pop}
Relaxation of the donor population for the electron transfer
model. The solid line shows the exact solution of the QME, the 
dashed line one arbitrary
trajectory, the dotted line an average over 500 trajectories.  }
\end{figure}

\begin{figure}
\includegraphics[width=8.cm]{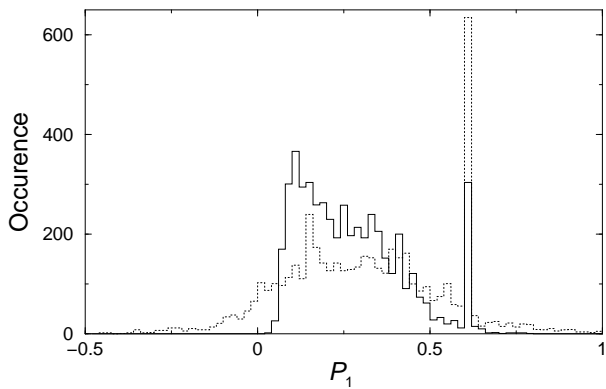}
\caption{
\label{fig:et-distrib}
Occurrence of the expectation values of the population on the
donor state produced by the new unraveling scheme for the Redfield QME
(dotted line) and the standard normalized jump method for the Lindblad QME
(solid line) at time $\omega t/(2\pi)=3$, both with 5000 trajectories.  }
\end{figure}

As a next test for the present quantum jump method we shall demonstrate the
stochastic unraveling of the Redfield QME \cite{may00,klei01}
\begin{eqnarray}
\dot{\rho} = - \frac{i}{\hbar} \left[ H_{\rm S},\rho \right]
+ \frac{1}{\hbar^2}
\Big\{
\left[ \Lambda\rho,K\phantom{^\dagger} \right]+
\left[ K,\rho \Lambda^{\dagger} \right]
\Big\}
\label{eq:redfield}
\end{eqnarray}
in which $\Lambda$ is the relaxation operator and $K$ the system part of
the system-bath interaction \cite{may00}.
Let us
consider a model for electron transfer in which the system includes a
single reaction coordinate with the Hamiltonian \cite{may00,klei01}
\begin{eqnarray}
H_{\rm S}=H_1|1\rangle\langle 1|+H_2|2\rangle\langle
2|+v_{12}(|1\rangle\langle 2|+|2\rangle\langle 1|)
\end{eqnarray}
where $H_1$ and $H_2$ are the Hamiltonians of two coupled harmonic
oscillators with frequency $\omega$. We choose a potential configuration in
the normal region with no barrier between the two harmonic potentials (change of free energy $\Delta
E=2\omega$, reorganization energy $\lambda=3\omega$) with inter-center
coupling $v_{12}=\omega$. The bath is described by a cut-off frequency
$\omega_c=\omega$ and temperature $kT=\omega/4$. The system-bath
interaction is characterized by the damping rate $\Gamma=\pi\eta/({\cal
  M}\exp(1))=\omega/10$ (see Ref.~\cite{klei01} for details).  After
rearrangement of Eq.~(\ref{eq:redfield}) one can easily identify the
operators involved in Eq.~(\ref{eq:qme}) ($M=1$):
$C_1=K$, $E_1=\Lambda$,  $A=-i H_{\rm S}-K\Lambda$.
A Gaussian wave packet located at the donor state $|1\rangle$ and having
energy slightly above the crossing of the harmonic potentials was chosen as
initial state. The numerical simulation for about 1000 trajectories
provides sufficiently converged and accurate results. Fig.~\ref{fig:et-pop}
shows the relaxation of the ensemble averaged donor population $\overline
{P_1}=\overline{\langle\psi|1\rangle\langle1|\phi\rangle}+
\overline{\langle\phi|1\rangle\langle1|\psi\rangle}$.  A widely discussed
property of the Redfield equation is that it does not conserve
positivity  \cite{may00}.  Although $\overline{P_1}$ is always
positive the tiny negative fraction in Fig.~\ref{fig:et-distrib} is
evidence for the existence of single realizations with negative $P_1$. The
simulation of the same system within the so-called diabatic-damping
approximation \cite{egor01,klei01} with a Lindblad QME by means of the
standard quantum jump method \cite{dali92,gard92,gisi92,garr94,plen98}
keeps all values of $P_1$ well confined between $0$ and $1$.

To summarize, a new method of stochastic unraveling of QMEs beyond the
Lindblad form is proposed and thus new large fields of application for
stochastic methods are opened.  This progress became possible with the use
of the wave--function pair in the doubled Hilbert space and the derivation
of stable, almost normalized SSEs. The efficiency is determined by the
behavior of the norm of every single trajectory. In this sense the jump
rates were used as parameters to influence the efficiency. Negative values
for the weight of single trajectories allow for the reconstruction of non
positive-semidefinite RDMs if required.  The method was successfully tested
for a simple electron transfer model and for Brownian motion and should
allow for better quantum dynamical simulation of large systems. It can also
unravel non-Markovian QMEs when they are in a time-local form like in the
time-convolutionless formalism \cite{cape94b} or in methods using auxiliary
density matrices to include the memory effects \cite{meie99} as well as
post-Markov master equations \cite{yu00}.


\end{document}